\documentclass[preprint,12pt]{elsarticle}
\usepackage[english]{babel}
\usepackage{url}
\usepackage{graphicx}
\usepackage{xtab}
\usepackage{listings}
\usepackage[noend]{algorithmic}
\usepackage{algorithm}
\usepackage{amssymb}
\usepackage{amsmath}
\usepackage{amsthm}
\usepackage{color}
\journal{Journal of Computational Physics}
\begin{document}
\begin{frontmatter}
\title{Reaction-diffusion model Monte Carlo simulations on the GPU}
\author[sch]{R.D.~Schram}
\ead{schram@lorentz.leidenuniv.nl}
\address[sch]{Instituut-Lorentz, Leiden University, P.O. Box 9506, 2300 RA  Leiden, The Netherlands}
\date{\today}

\begin{abstract}
 We created an efficient algorithm suitable for graphics processing units (GPUs) to perform Monte Carlo simulations of a subset of reaction-diffusion models. The algorithm uses techniques that are specific to GPU programming, and combines these with the multispin technique known from CPU programming to create one of the fastest algorithms for reaction-diffusion models. As an example, the algorithm is applied to the pair contact process with diffusion (PCPD). Compared to a simple algorithm on the CPU, our GPU algorithm is approximately 4000 times faster. If we compare the performance of the GPU algorithm, between the GPU and CPU, we find a speed-up of about 130x.
\end{abstract}

\end{frontmatter}
\section{Introduction}

One-dimensional reaction-diffusion models have received interest over the past decades. In non-equilibrium statistical mechanics these include directed percolation (DP), pair contact process with diffusion (PCPD) and triple contact process with diffusion (TCPD). The interest in these models is mostly whether we can classify the models into one, or more universality classes. Grassberger \cite{grass} and Janssen \cite{janssen} conjectured  that all systems with a single order parameter and a single absorbing state will belong to the universality class of the Directed Percolation model. The order parameter of the reaction-diffusion models considered here is simply the density $\rho$ of the system. 

Monte Carlo simulations are very useful in this area of research, because the critical behaviour of these simple models is not understood very well theoretically. The Monte Carlo approach was extensively used on the PCPD model, with varying degrees of success. As shown in \cite{ref:hinrichsen}, there are strong finite-time corrections. Thus, both long time-series and good statistics are mandatory for a proper analysis of the critical behaviour of the system. Acquiring sufficient amounts of data is a very time consuming effort. Thus, we used the parallel processing power of the GPU and clever processing techniques to achieve this in Ref. \cite{schram_pcpd}. This paper will concentrate on the implementation and benchmarks of this algorithm. 

This paper uses the PCPD model as an example model that can be simulated using the described algorithm. It is very easy to expand the algorithm to other related models, such as the DP, TCPD and QCDP (Quadruplet Contact Process with Diffusion) models. In fact, the algorithm has already been applied to the TCPD and QCPD models, and analysis of these models will hopefully be published in the near future.

The graphics processing unit (GPU) has been utilized in the last decade for the purpose of gathering high quality data in computational physics: \cite{ref:preis09} (Ising model), \cite{ref:anderson08} (Molecular dynamics) and \cite{nedelcu11} (bond fluctuation model). This paper combines GPU programming with a clever programming technique called multispin programming, that originates from general purpose CPU programming. It has been used for the PCPD problem in Ref. \cite{small2008}. A more thorough explanation of this technique is found in \cite{book_bar}. The relevant parts for our purposes are explained in this paper. 


\section{Pair contact process with diffusion}

The PCPD model is a 1+1 dimensional problem from non-equilibrium statistical mechanics. In this model, particles on a lattice can interact in two ways when they are adjacent to each other: both particles can annihilate, or a new particle can be produced adjacent to either particle. Additionally, single particles can diffuse on the lattice. More formally, the reactions are described by the following notation, where A denotes a particle and 0 denotes an empty site on the lattice:

\begin{eqnarray}
\begin{array}{ccc}
\left\{
\begin{array}{ccc}
AA0 & \rightarrow & AAA \\
0AA & \rightarrow & AAA
\end{array} \right.
&{\rm each\,\,with\,\, rate}&
\frac{(1-p)(1-d)} 2 \label{2Ato3A} \\
AA~ \rightarrow ~ 00
&{\rm with \,\,rate}& p\,(1-d) \label{2Ato0}\\
A0  \leftrightarrow  0A &{\rm with \,\,rate}&  d \label{A0to0A}
\end{array}
\label{eq:rates}
\end{eqnarray}

\begin{figure}\centering
\includegraphics[width=10cm]{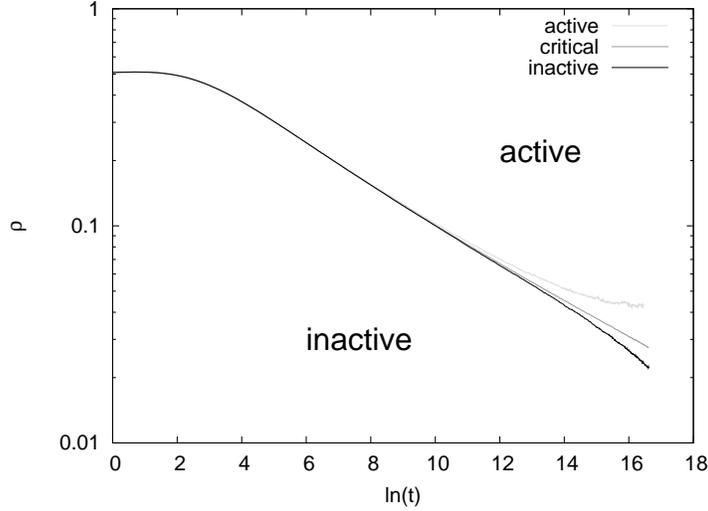}
\caption{The density of the system as a function of time for the PCPD model, with three different $p$ values: one in the active phase, one close to critical, and one in the inactive phase.}
\label{fig:phase}
\end{figure}

There exist three regimes of the system. A schematic overview is shown in figure \ref{fig:phase}. For a low annihilation rate $p<p_c$, the system will maintain a constant density with very high probability. This is called the active phase. On the other hand, in the case of a high annihilation rate $p>p_c$, the particles in the system will annihilate quickly, and the system is in the inactive phase. On the boundary at the critical annihilation rate $p=p_c$, length and time scales are expected to diverge in a power law fashion, similar to equilibrium phase transitions. The actual value for $p_c$ depends on the diffusion coefficient $d$. For $p$ values far from the critical value $p_c$, the behaviour of the system is well described by mean-field theory. Thus, our interest lies in determining the critical behaviour at or close to $p_c$.

A sequential Monte Carlo algorithm of the PCPD model is shown in algorithm \ref{alg:seq}. It is not optimized for speed, because for instance the number of calls to the random number generator (RNG) can be reduced to only one call. The reason for writing the algorithm this way is that it compares more easily to the GPU algorithm. 

\begin{algorithm}[tb]
\caption{Sequential PCPD algorithm}
\begin{algorithmic}[1]\label{alg:seq}

	\STATE \textbf{Input: } configuration $q$ at time $t$.
	\STATE \textbf{Output: } configuration $q$ at time $t+1/L$.

	\medskip
	\STATE \textbf{PCPD\_step\_seq}($q$):
	\STATE \COMMENT{pick a site $i$ on the lattice}
	\STATE $i := \lfloor L\cdot RNG() \rfloor$ \COMMENT{RNG is uniform in $[0,1)$}
	\STATE $r_1 := RNG()$, $r_2 := RNG()$, $r_3 := RNG()$
	\medskip
	\IF{$r_1<d$}
		\STATE Swap($q_i$, $q_{i+1}$) \COMMENT{Diffusion}
	\ELSIF{ $q_i=1$ and $q_{i+1}=1$}
		\IF{$r_2<p$}
			\STATE $q_i := 0,$ $q_{i+1} := 0$ \COMMENT{Annihilation}
		\ELSE
			\IF{$r_3<1/2$}
				\STATE $q_{i-1} := 1$ \COMMENT{Fission low}
			\ELSE
				\STATE $q_{i+2} := 1$ \COMMENT{Fission high}
			\ENDIF
		\ENDIF
	\ENDIF

\end{algorithmic}
\end{algorithm}

\section{GPU programming}

Some algorithms are more easily ported to a GPU architecture than others. It is important to understand the strengths of the GPU to see whether it is worth the effort. First we will give the conditions under which a GPU program can thrive. Our implementation uses the OpenCL \cite{opencl} framework for parallel computing. Therefore, OpenCL terminology will be used to describe the basics of GPU architecture. The difference between the two GPU architectures of the main GPU manufacturers AMD and NVIDIA is far smaller than difference between a CPU and a GPU. For our purpose the difference is small enough, that it does not need a different implementation. Technical specifications of the GPU architectures can be found in \cite{nvidia} (NVIDIA) and \cite{amd} (AMD).

The first and most important condition is the amount of parallelism that can be obtained in the algorithm. For the current generation GPUs, the number of threads should ideally be of an order 10,000 or higher per GPU. Future GPUs will be even more parallel, and this requirement will further increase. Generally speaking, for Monte Carlo simulations there are two (easy) methods to increase parallelism: increase the system size and let more threads work on one system, or increase the number of statistically independent systems and run them in parallel. The second option is easy to implement, but often for the desired system size the simulation will either consume too much memory, or time scales that can be achieved are too low, which is especially important in the case of non-equilibrium models such as PCPD.

Having more threads working on the same system typically relieves these problems, but it is harder to program and it adds synchronization cost to the total computation cost, which can be severe if all threads on the GPU are synchronized.

On a GPU the \textit{workitems}, which are called threads on a CPU, are divided into workgroups. The programmer can decide on the size of a workgroup, up to a maximum that depends on the hardware. Synchronization and communication between workitems within a workgroup is cheap on the GPU, compared to synchronization and communication between two workitems that belong to a different workgroup. To reduce communication overhead, in our implementation each workgroup simulates its own independent system with a different seed of the random number generator (RNG). 

Attaining a high memory bandwidth utilization is another aspect of an efficient GPU algorithm. The memory bandwidth of a current-gen high-end GPU (AMD Radeon HD7970, 260 GB/s) is about 5 times higher than that of a current-gen high-end CPU (Intel Core i7 3960X, 51 GB/s). However, the memory access pattern is more important on a GPU; a bad access pattern such as random access will hurt the performance on the GPU more than on the CPU. Sequential access gives the best performance, but the PCPD problem requires us to randomly select a lattice site, which is an inefficient memory access pattern. We circumvent this by slightly modifying the PCPD algorithm. The lattice is divided among the workitems into parts of equal sizes. Then as a compromise, we modify the PCPD algorithm such that all workitems access the same local site in their part of the lattice, where the local site is the site relative to the appointed part of the lattice. The local site for all the workitems is still randomly chosen. Local sites with the same number can be placed sequentially, and the random access pattern becomes at least partially sequential. A more detailed description of how the lattice sites are ordered in memory will be given in section \ref{sec:multi}.

Since the reaction-diffusion models are simulated using Monte Carlo algorithms, it is important to carefully choose the random number generator (RNG). One aspect that determines the quality of a good RNG is the number of bits that is used to store the current state of the RNG. There is a comparatively large, but still limited amount of general purpose registers (GPRs) available on the GPU. Access to these registers is in principle without delay, barring some restrictions that are handled by the compiler. To optimize for performance we chose the combined Tausworthe RNG with the seeds from Ref. \cite{lecuyer99}. It has a 128-bit state storage, which can easily fit inside the GPRs. To verify that this does not deteriorate the results, we used the WELL512 RNG with a 512 bit state space, to check whether the results are influenced by the small state space. We found that this is not the case.

The final important consideration that we want to highlight is the use of branching on the GPU. Workgroups on the GPUs of both NVIDIA and AMD are divided into (smaller) entities called warps by NVIDIA and wavefronts by AMD. These warps/wavefronts are executed in SIMD (Single Instruction Multiple Data) fashion. This means that each workitem in a warp/wavefront executes the exact same instructions. Instructions within a branch are executed by all workitems within a warp/wavefront if at least one workitems takes that particular branch. In addition to that, older AMD GPUs (prior to the Core Gen Next architecture) have a high latency penalty when encountering a branch, which is at least $\sim 50$ clock cycles. In our implementation all branching operations in the simulation part of the program are replaced by bit-operations. 

\section{The multispin GPU program} \label{sec:multi}

Multispin programming is a technique that enables parallelism within one thread. It can be used when the number of bits necessary to store one data element is smaller than the register size of the underlying hardware. For example, modern 64-bit CPUs have 64-bit registers (not counting SSE, AVX extensions), whereas current GPUs work fastest on 32-bit integers/floating point numbers. For the reaction-diffusion models that we consider, a lattice site can either be occupied, denoted by 1, or empty, denoted by a 0. Thus, to store the state of one lattice site, we only need one bit of storage per site. In the original multispin technique, parallelism is obtained by simulating $N_r$ lattices in parallel, where $N_r$ is the number of bits in the GPR, depending on hardware. The sequential program is translated to use bit operations, instead of branching statements. Finally, the lattices are decoupled in some way, to ensure that after sufficient time, the $N_r$ lattices are statistically independent. 

We opted for a slightly different approach. The problem with the classical approach to multispin programming in the case of GPUs is twofold. Firstly, the multispin technique does not make time steps go faster, but only improves the statistics of the results. Thus, to obtain long time series, we still need to wait a long time, although the statistics are much better than on a CPU, because of the number of lattices used. Perhaps more importantly, the necessary amount of memory grows very large. In our simulations we used a lattice size of $L=2^{18}=262144$. With $500$ workgroups, the total amount of memory used would be about $2$ GB of memory. Although professional Tesla, or FireGL cards usually contain this amount of memory, it does reach the limit of most consumer cards (which are used extensively in simulations because of the price difference), and it does not leave much room for increasing the lattice size. 

Instead, we connected the $N_r$ lattices of the classical approach, to form one larger lattice. A concern that may arise, is that the correlation between the sites within one GPR (that of one workitem) will propagate and bias the results. We found that this is indeed the case if sites in the same GPR are close to each other, but if they are the maximum distance apart ($L/N_r$), we found no bias, as a consequence of this trick to increase the lattice size. Additionally, we added extra decoupling between sites in the same GPR, which will be discussed in section \ref{sec:dif}.

\begin{figure}\centering
\includegraphics[width=10cm]{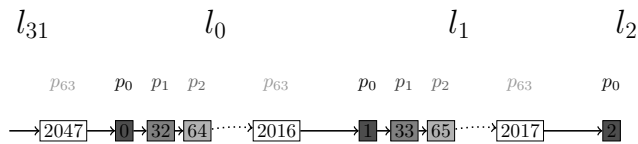}
\caption{A schematic picture of the memory layout of the multispin GPU algorithm. The configuration shown in this figure has a work-group size of 64. The number of local elements is 32. Consequently, the lattice size is $L=32\cdot 32\cdot64 = 65536$. The work-item number is given by $p_{i}$. The local memory index is denoted by $l_i$. The numbers in the boxes give the global lattice site number, which means that $s=1$ is a neighbor of $s=2$, etc. The arrows indicate lattice sites that are next to each other in the global memory of the GPU. The vertical columns are not drawn, because they are the same as in figure \ref{fig:mem_dist}, i.e. only the first bit of the double word is shown. For example, global lattice site 2048 is contained in the same double word as 0.}
\label{fig:mem_conf}
\end{figure}

\begin{figure}\centering
\includegraphics[width=10cm]{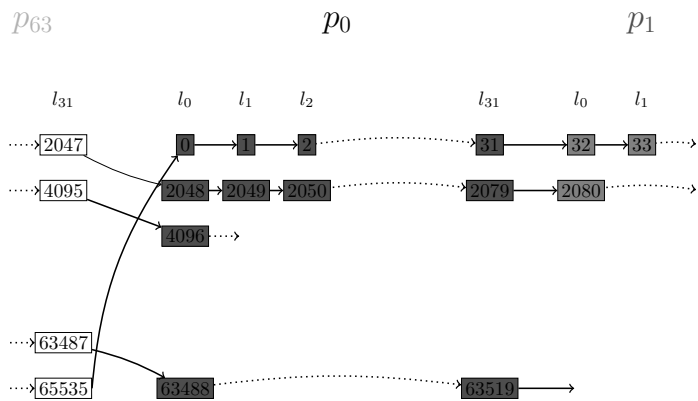}
\caption{A schematic figure showing the distribution of the sites of the lattice among the work-items. The work-item number is given by $p_{i}$. The local memory index is denoted by $l_i$. The numbers in the boxes are global site numbers, where $s=1$ is a neighbor of $s=2$, etc. The arrows indicate consecutive lattice sites. A vertical column is stored in one double word, and this is where multispin coding is used.}
\label{fig:mem_dist}
\end{figure}

The placement of the sites in memory is shown in figures \ref{fig:mem_conf} and \ref{fig:mem_dist}. This memory lay-out is optimized for performance. At each step of the simulation of the reaction-diffusion model, the GPU program loads a number of consecutive sites into the GPRs of each workitem. The sites loaded from the global memory are those that are needed for computation of the Monte Carlo step, and those that can potentially be modified. Thus, for the PCPD problem, four sites loaded into the GPRs (the pair, one to the left of the pair and one to the right). For a equivalent definition of the TCPD problem, five sites would have to be retrieved. The workgroup size is chosen to be a power of 2, such that the memory loads are $4 \cdot S_w$ byte aligned, with $S_w$ the size of the workgroup. According to a technical document of AMD \cite{amd}, the memory channel width of AMD GPUs is 256 byte, and we expect at least decent performance, using a workgroup size of $64$. 

The distribution of work among the workitems in combination with the memory lay-out ensures that there is no communication needed between workitems: the sites on which the workitems work are always far enough ($L/N_r$) apart such that they are independent within a step. The sites are stored in the (shared) global memory, which also prevents the need for communication, as opposed to using the GPRs. Between computation steps, however, it is necessary to synchronize the workitems within a workgroup to avoid race conditions on the global memory. If the workgroup size is small enough, this synchronization is a intrinsic property of the SIMD array of the GPU and in that case it is cost-free (64 for AMD GPUS, 32 for NVIDIA GPUs). 

\subsection{Implementation details of the PCPD algorithm}

\begin{algorithm}[tb]
	
	\caption{GPU multispin PCPD algorithm}
	\begin{algorithmic}[1]\label{alg:pcpd_gpu}
	
	\STATE \textbf{Input: } Local lattice $l$ at time $t$, site(s) $i$, percolation probability $p$.
	\STATE \textbf{Output: } Local lattice $l$ at time $t+1/32$.
	\STATE \textbf{Call: PCPD\_GPU\_multi}($L$, $i$):
	
	\medskip
	\STATE \COMMENT{Diffusion, $d=0.5$}
	\STATE $\mathrm{dMask} := \mathrm{RNG()}$
	\STATE $\mathrm{temp} := l[i]$ \label{all:drest}
	\STATE $l[i] := (\mathrm{dMask}\land l[i+1]) \lor ((\lnot \mathrm{dMask}) \land l[i])$
	\STATE $l[i+1] := (\mathrm{dMask}\land\mathrm{temp}) \lor ((\lnot \mathrm{dMask}) \land l[i+1])$ \label{all:drestend}
	\medskip
	\STATE \COMMENT{pMask has the $j$-th bit set to 1 iff the $j$-th bit in both $l[i]$ and $l[i+1]$ are 1}
	\STATE $\mathrm{pMask} := l[i]\land l[i+1]$ 
	\medskip
	\STATE \COMMENT{Annihilation, $p<0.25$}
	\STATE $\mathrm{aMask} :=$ RNG() $\land$ RNG() \label{all:dec}
	\STATE $\mathrm{aMask} := (\lnot \mathrm{dMask}) \land \mathrm{aMask} \land \mathrm{pMask} \land  (({\rm RNG}()< 4 p)?1:0)$ \label{all:glob}
	\STATE $l[i] := l[i] \land \mathrm{aMask}$
	\STATE $l[i+1] := l[i+1] \land \mathrm{aMask}$
	\medskip
	\STATE \COMMENT{Fission high}
	\STATE $\mathrm{fMask} := \mathrm{pMask} \land  (\lnot \mathrm{dMask}) \land (\lnot \mathrm{aMask}) \land \mathrm{RNG()}$ \label{all:fis_start}
	\STATE $l[i+2] := l[i+2] \lor \mathrm{fMask}$
	\medskip
	\STATE \COMMENT{Fission low}
	\STATE $\mathrm{fMask} := \mathrm{pMask} \land  (\lnot \mathrm{dMask}) \land (\lnot \mathrm{aMask}) \land (\lnot \mathrm{fMask})$
	\STATE $l[i-1] := l[i+2] \lor \mathrm{fMask}$ \label{all:fis_end}
  \end{algorithmic}
\end{algorithm}


To keep the pseudocode compact and more easy to follow, we assume that the required sites are already loaded from the global memory into the GPRs, and that the necessary bit-rotations to connect the multi-spin lattices are already done. 

The PCPD-specific core of the algorithm is given in Algorithm \ref{alg:pcpd_gpu}. The bitwise operations used are the \textbf{or} ($\lor$), \textbf{and} ($\land$) and \textbf{not} ($\lnot$) logical operators. Except for the comparison in line \ref{all:glob}, the pseudocode can be read as if it were only working on one bit. In the processing unit of the GPU these bit-wise operations are done on all $N_r$ bits simultaneously. Line \ref{all:glob} is an exception: the comparison ${\rm RNG}()< 4 p$ is not done bit-wise. Instead, it gives either 1 or 0 equally for all $N_r$ bits. The reason for this will be given in section \ref{sec:anni}.

\subsection{diffusion} \label{sec:dif}

In algorithm \ref{alg:pcpd_gpu} the diffusion coefficient $d$ is set to $0.5$. In contrast to a straightforward PCPD implementation, the value for $d$ is hard coded for efficiency. The advantage lies in  the computational effort to create a bit that is one with probability $d$. For $d=0.5$ the output of the function RNG$()$ (with a correct implementation of the RNG) is 1 with probability $0.5$, and 0 otherwise. To obtain other values for $d$ \textbf{or} and \textbf{and} operators are used. For example to create bits that are 1 with probability $0.25$ the \textbf{and} operator is used between two random numbers: $\mathrm{dMask} = \textrm{RNG}() \land \mathrm{RNG}()$. This way we can generate $2^{N_g}-1$ different values of $d$ using at most $N_g$ calls to the RNG ($d=0$ is not counted). The universal behaviour of the system is believed to independent of $d$, and therefore the limited choice of $d$ is not an issue, as long as we can select values for $d$ that differ sufficiently.

In lines \ref{all:drest}-\ref{all:drestend} the diffusion step is done by swapping the bits of sites $l[i]$ and $l[i+1]$, if the bit in $\mathrm{dMask}$ is 1.

\subsection{Annihilation} \label{sec:anni}

In the annihilation step we are confronted with a similar problem as in the diffusion step. We need to create a set of $N_r$ bits that are one with probability $p$. We could do this analogously with the diffusion step, but in this case an arbitrary value for $p$ is not useful, because $p$ needs to be close to $p_c$. In our simulations in Ref. we found $p_c$ with an error of approximately $2 \cdot 10^{-6}$, which means that we would need at about $18$ random numbers to have multiple values of $p$ close enough to $p_c$ ($2^{-18} \approx 4\cdot10^{-6}$). The final PCPD program needs 5 or 6 random numbers in total, depending on $d$. Thus, it would be a bottleneck if we would use the same procedure for the annihilation step. Instead, all $N_r$ bits are set to 1 with probability $p$ and all 0 otherwise. This gives each bit the right probability to be 1, but the bits are fully correlated. With the described memory lay-out in section \ref{sec:multi} we found the bias due to the correlation to be smaller than the noise of our data, but to be safe we added a decorrelation step to our algorithm. 

We found in Ref. \cite{schram_pcpd} that $p_c<1/4$ for $d=0.25$, $d=0.5$ and $d=0.75$. In line \ref{all:dec}, $N_g$ bits are generated independently of each other being 1 with probability $0.25$. Then in the next line we apply the \textbf{AND} operator between these bits and the fully correlated bits that are one with probability $4 p$. The result is a partially correlated bit with probability $p$ to be one. In the case that we would not have the property $p<1/4$, we can still do a similar procedure. For example, in the case of the bound $p<3/4$, we would generate uncorrelated bits with a probability of $3/4$ to be 1, and use the \textbf{AND} operation with correlated bits that are one with probability $4p/3$. The quality of the decorrelation procedure depends on the tightness of the bound we use.

The last two lines clear the two sites $l[i]$ and $l[i+1]$ under the following conditions:

\begin{enumerate}
 \item $l[i] = l[i+1] = 1$
 \item No diffusion this step
 \item We drew 1, with a probability $p$
\end{enumerate}

\subsection{fission}

Lines \ref{all:fis_start}-\ref{all:fis_end} constitute the fission step. By the definition given in eq. (\ref{eq:rates}), the conditional probability to attempt a fission step, given that no diffusion or annihilation step was attempted, is 1. Therefore, only one random number is necessary to distinguish between a new particle to the left and to the right of the pair. 

\section{Benchmarks}

\subsection{Test setup}

We use the ``Little Green Machine'' for our main benchmarking tests. A node consists of two NVIDIA Geforce GTX 580 GPUs, an Intel Xeon E5620 (@2.4 Ghz) CPU, and 24 GB of memory. As the OpenCL library we use the NVIDIA Cuda toolchain, version 4.0. The program was compiled using version 4.1.2 of the gcc compiler, with optimization flags -O2.

Additionally to this setup, we also use a PC, consisting of an ATI Radeon HD5850 GPU, an Intel Core 2 Duo E8400 (@3.0 Ghz), and 4 GB of memory. 

\subsection{Performance}

\begin{figure}\centering
\includegraphics[width=10cm]{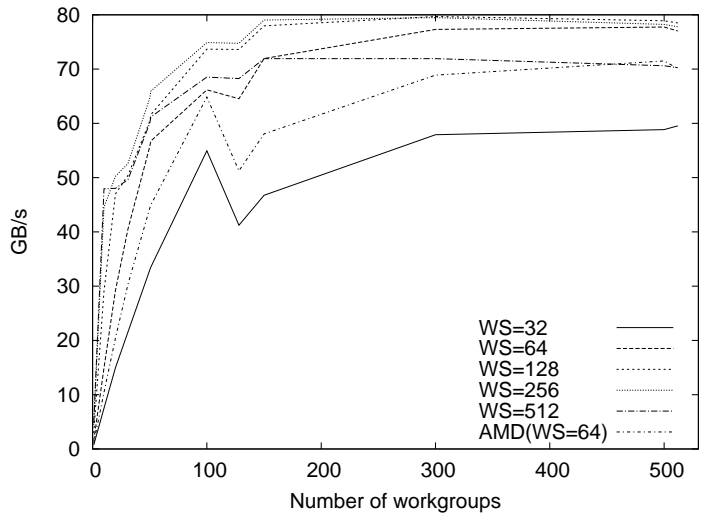}
\caption{Performance of the GPU PCPD algorithm, depending on the number of workgroups and the size of the workgroups. The performance is measured in the effective bandwidth the algorithm achieves. For the PCPD problem, the number of bytes transfered per second is equal to the number of moves per second. The number of workgroups is equal to the number of parallel lattices on the GPU. The smallest workgroup (WS=32) has the worst performance. However, it has theoretically the most numerically most accurate results. Thus, for safety for our analysis \cite{schram_pcpd} we used a workgroup size of 32. The fact that the graph is not smooth, is not due to statistical fluctuation. The error bars on the curves is much less than 1 GB/s. Instead, the reason is that the performance does not depend smoothly on the number of workgroups, because the number of computational units is an integer. We will not discuss more deeply why some workgroup size is better performing than others.}
\label{fig:perf}
\end{figure}

The performance of the GPU PCPD algorithm is measured in the number of moves done per second globally on the GPU, including multispin. In the PCPD case the program reads 4 bits, and writes back 4 bits to the lattice, per single move. Thus the total (read/write combined) bandwidth in bytes/s (B/s) of the PCPD program is equal to the number of moves/s. The performance of the PCPD GPU program is shown in figure \ref{fig:perf}, for different GPUs and a CPU. 

Our conventional CPU program does roughly $20 \cdot 10^6$ moves per second, which means that the GPU program is about 4000 times faster. The same GPU program run on the CPU performs $615 \cdot 10^6$ moves per second, which is still 130 times slower than the same algorithm run on a GPU. Using both cores, the algorithm achieves almost perfect scaling with $1200 \cdot 10^6$ moves per second. Thus, we conclude that this algorithm is very well tailored for the GPU. We believe that this is due to the fact that the algorithm is still computationally bound, where the GPU has the largest architectural advantage. Also, the GPU algorithm on the CPU has the disadvantage that the memory access pattern is more simlar to a random access pattern.

The performance depends on the size of the workgroups on the GPU. The advantage of a smaller workgroup is that synchronization is less expensive. Also, the number of workgroups that fits on a compute unit/SIMD array concurrently is larger because the stacking of workgroups is less coarse grained. In our implementation the number of bytes read sequentially is equal to $4 S_w$. The more bytes that are read sequentially, the better the global access pattern, which can improve performance. With a workgroup size of 32 we are likely seeing this effect: even with the largest number of workgroups, it is still 20\% slower than the simulations with larger workgroup sizes. Since less workitems per workgroup makes the simulation more safe with regards to finite size effects, the best choice is probably either $WS=64$ or $WS=128$. 

To test the numerical dependence of the density $\rho(t)$ on the workgroup size and the number of decorrelation bits, we have used different configurations of those. Only with the largest workgroup size (512) there is a significant deviation from the other curves. Adding decorrelation bits does not help to improve the numerical result. This means that the cause of the deviation is the choice of sites, which is correlated between the workitems (and which is not influenced by decorrelation bits). On the other hand, a workgroup size of $128$ is indistinguishable from runs with $S_w=32$. Thus, we conclude that our data with $S_w=32$ is almost certainly not significantly biased by the finite lattice size. 

\subsection{Analysis}

The analysis was done in a previous paper \cite{schram_pcpd}. Here, we will present some analysis of the data to show the necessity of a fast algorithm. 

At the critical value $p_c$, the density $\rho$ of the system is expected to decrease in a power-law fashion. However, for the PCPD problem we found that there are strong corrections to scaling, which we believe to be of the following form:

\begin{equation}
 \rho(t) = c_1 t^{-\delta} + c_2 t^{-2 \delta} + \ldots.
\end{equation}

\begin{figure}\centering
\includegraphics[width=10cm]{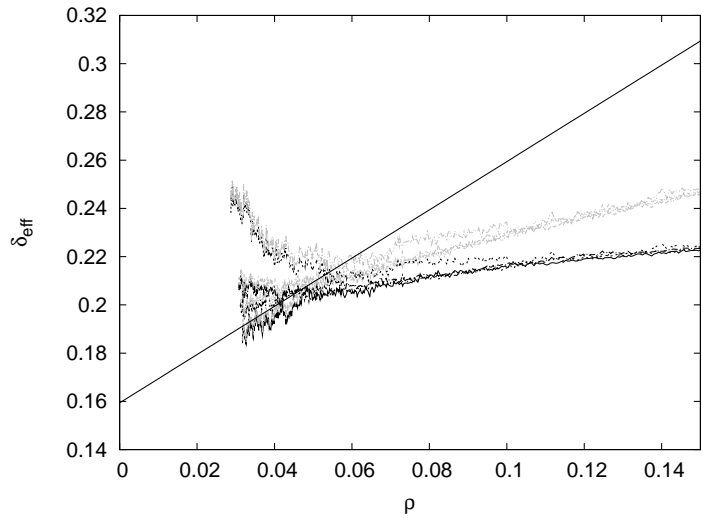}
\caption{The effective exponent $\delta_{\rm eff}$ for both the single particle density (black lines) and the pair density (gray lines), as a function of their respective densities. The diffusion constant $d$ is equal to $0.25$. The values for $p$ was chosen close to the critical point, that was estimated to be $p_c = 0.125141(2)$. The lattice size was chosen to be $L=2^{18} = 262144$. The values for $p$ are 0.12514 ($N=2000$), 0.125142 ($N=3000$), 0.125145 ($N=3000$) and 0.125155 ($N=1000$), with $N$ the number of runs.}
\label{fig:delta_0.25}
\end{figure}

\begin{figure}\centering
\includegraphics[width=10cm]{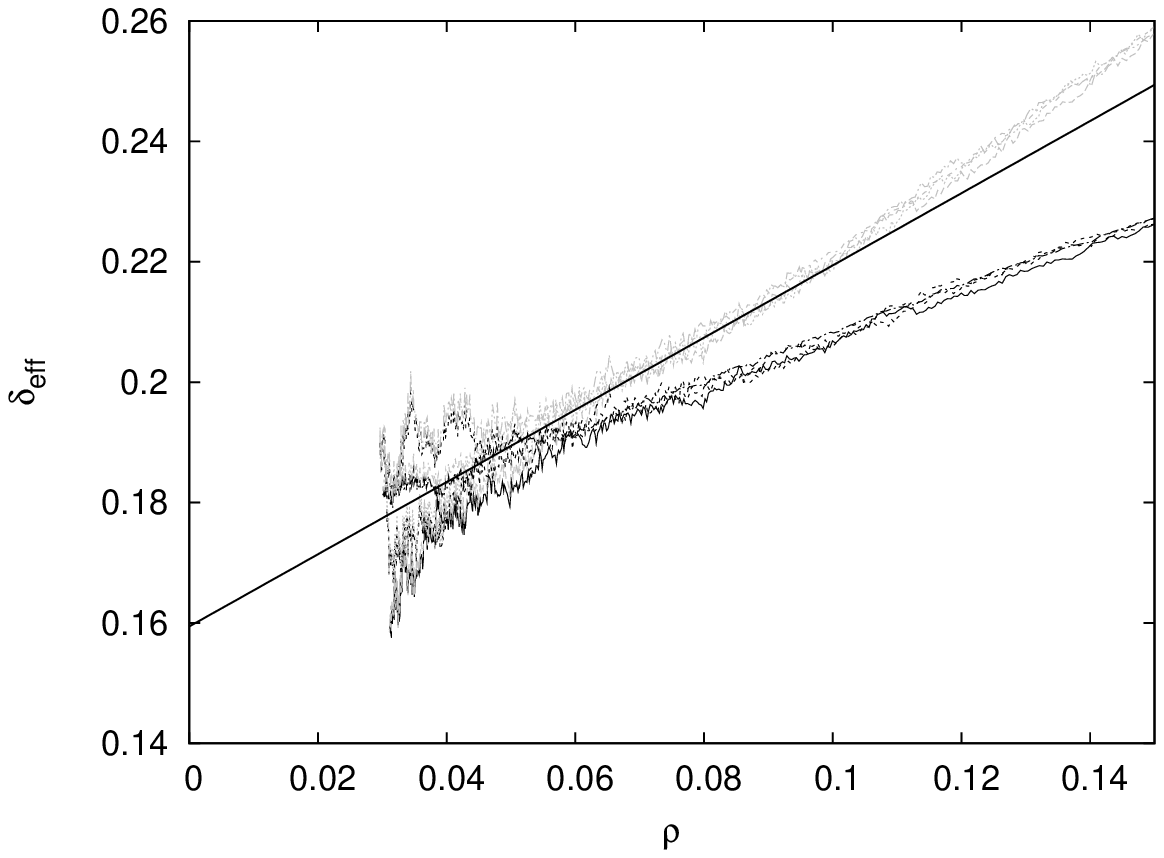}
\caption{The effective exponent $\delta_{\rm eff}$ for both the single particle density (black lines) and the pair density (gray lines), as a function of their respective densities. The diffusion constant is equal to $0.75$. The values for $p$ was chosen close to the critical point, that was estimated to be $p_c = 0.191789(2)$. The lattice size was chosen to be $L=2^{18} = 262144$. The values for $p$ are 0.19178 ($N=2000$), 0.191785 ($N=2000$), 0.19179 ($N=8000$) and 0.191795 ($N=2000$), with $N$ the number of runs.}
\label{fig:delta_0.75}
\end{figure}

As shown in the analysis paper, under this assumption, asymptotically a straight line is obtained by plotting $\delta_{\rm eff}$ as a function of $\rho$. This is one in figures \ref{fig:delta_0.25} and \ref{fig:delta_0.75} for diffusion coefficients $d$ equal to 0.25 and 0.75 respectively. The figure with $d=0.5$ was shown in the analysis article. The particle density and the pair density are expected to converge to the same value of $\delta$ at $\rho=0$ (or equivalently $t \rightarrow \infty$). The pair density and the particle density at high densities (low $t$) have slopes that when extrapolated, cross each other at some finite density. Thus, at these time scales, it is not possible to extract the universal exponent $\delta$ by extrapolation. The data after the virtual crossover point is more useful in this regard, but even with the amount of data gathered, this part is still quite noisy and short. This shows the necessity of a fast algorithm as presented in this paper.

\section{Conclusion and discussion}

We have addressed concerns about the numerical accuracy, as a result of an insufficiently large lattice, in combination with a large workgroup size. We conclude that the effective lattice size does indeed decrease with an increasing workgroup size. Thus, the effective lattice size will be smaller than the number of sites in the lattice. This means that the lattice size has to be bigger than in an ordinary CPU simulation. However, the simulations presented here were not limited by the amount of memory, and for workgroup sizes smaller than $128$, we did not observe any deviation at all, with a lattice size of $2^{18}$, and $t<1.5 \cdot 10^{7}$.

The GPU reaction-diffusion algorithm presented here is much faster than traditional multispin methods, which are among the fastest CPU methods.  With our algorithm, one GPU is roughly equivalent to $130$ CPU cores. A multispin program specifically written for the CPU will be faster, but it is unlikely to be much more than twice as fast (most can be gained from going from 32-bit integers to 64-bit integers), which still leaves a factor of roughly $65$. Thus, it greatly sped up the simulations used to analyze the PCPD model, which otherwise would have taken an immense amount of CPU hours. 

The algorithm is easy to extend to other related models, and is already being used for that problem. Hopefully this will shed more light on the issue of universality in non-equilibrium statistical mechanics models.

\section{Acknowledgements}

Computing time on the ``Little Green Machine", which is funded by the Dutch
agency NWO, is acknowledged.

\bibliographystyle{ieeetr}
\bibliography{imp_paper}

\begin{thebibliography}{10}

\bibitem{grass}
P.~grassberger, ``On phase transitions in schl\"ogl's second model,'' {\em Z.
  Phys. B}, vol.~47, no.~4, pp.~365--374, 1982.

\bibitem{janssen}
H.~K. Janssen, ``On the nonequilibrium phase transition in reaction-diffusion
  systems with an absorbing stationary state,'' {\em Z. Phys. B}, vol.~42,
  no.~2, pp.~151--154, 1981.

\bibitem{ref:hinrichsen}
H.~Hinrichsen, ``The phase transition of the diffusive pair contact process
  revisited,'' {\em Phys. A: Stat. Mech. and its Applications}, vol.~361,
  no.~2, pp.~457--462, 2006.

\bibitem{schram_pcpd}
R.~D. Schram and G.~T. Barkema, ``Critical exponents of the pair contact
  process with diffusion,'' {\em J. Stat. Mech.}, vol.~2012, no.~3, p.~3009,
  2012.

\bibitem{ref:preis09}
T.~Preis, P.~Virnau, W.~Paul, and J.~J. Schneider, ``Gpu accelerated monte
  carlo simulation of the 2d and 3d ising model,'' {\em Journal of
  Computational Physics}, vol.~228, no.~12, pp.~4468--4477, 2009.

\bibitem{ref:anderson08}
J.~A. Anderson, C.~D. Lorenz, and A.~Travesset, ``General purpose molecular
  dynamics simulations fully implemented on graphics units,'' {\em Journal of
  Computational Physics}, vol.~227, no.~10, pp.~5342--5359, 2007.

\bibitem{nedelcu11}
S.~Nedelcu, M.~Werner, M.~lang, and J.-U. Sommer, ``Implementations of the bond
  fluctuation model,'' {\em Journal of Computational Physics}, vol.~231, no.~7,
  pp.~2811--2824, 2011.

\bibitem{small2008}
F.~Smallenburg and G.~T. Barkema, ``Universality class of the pair contact
  process with diffusion,'' {\em Phys. Rev. E}, vol.~78, pp.~31129--31136,
  2008.

\bibitem{book_bar}
M.~Newman and G.~Barkema, {\em Monte Carlo Methods in Statistical Physics}.
\newblock Clarendon Press, Oxford, 1999.

\bibitem{opencl}
{Khronos Group}, {\em OpenCL specifications}, 2012.

\bibitem{nvidia}
{NVidia Corp.}, {\em Cuda programming guide}, 2012.

\bibitem{amd}
{Advanced Microdevices Inc.}, {\em AMD Accelerated Parallel Processing, OpenCL
  programming guide}, 2012.

\bibitem{lecuyer99}
P.~L'Ecuyer, ``Tables of maximally equidistributed combined lfsr generators,''
  {\em Math. Comp.}, vol.~68, pp.~261--269, 1999.

\end{thebibliography}

\end{document}